# Oxygen Vacancies in N doped anatase TiO$_2$:  Experiment and first-principles calculations

Abdul K. Rumaiz[1], J.C. Woicik[2], E. Cockayne[2], H.Y. Lin[3], G. Hassnain Jaffari[4] and S. I. Shah[3, 4]

1. National Synchrotron Light Source, Brookhaven National Laboratory, Upton, NY 11973, USA.
2. National Institute of Standards and Technology, Gaithersburg, MD 20899, USA.
3. Department of Materials Science & Engineering, University of Delaware, Newark, DE 19716, USA.
4. Department of Physics & Astronomy, University of Delaware, Newark, DE 19716, USA.

We have determined the electronic and atomic structure of N doped TiO$_2$ using a combination of hard x-ray photoelectron spectroscopy (HAXPES) and first-principles density functional theory calculations.  Our results reveal that N doping of TiO$_2$ leads to the formation of oxygen vacancies and the combination of both N impurity and oxygen vacancies accounts for the observed visible light catalytic behavior of N doped TiO$_2$.

Titanium dioxide (TiO$_2$) is widely studied for applications such as photocatalysis [1, 2], photovoltaics [3], and dilute magnetic semiconductor [4, 5]. One of the most promising applications of TiO$_2$ is its catalytic capability of splitting water into oxygen and hydrogen [6]. However, due to its large band gap (3 and 3.2 eV for rutile and anatase, respectively), pure TiO$_2$ is only activated by ultraviolet light, lowering the efficiency of its catalytic process. Doping of TiO$_2$ can introduce energy levels in the band gap [7, 8], effectively tailoring its electronic structure to absorb light in the visible region. Recently N doped TiO$_2$ has been synthesized and its catalytic properties demonstrated with visible light [9, 10]. This success, however, is riddled with controversies concerning the electronic structure of N doped $TiO_2$. Asahi et al [11] have proposed a band gap narrowing due to the lower binding energy of the N *2p* relative to the O *2p* levels, while Valentin et al [12] have proposed a localized dopant N *2p* state just above the O *2p* valence band maximum. While optical absorption measurements of N doped $TiO_2$ have shown band gap reduction [10, 13, 14], the corresponding change in the valence band has not been observed by photoemission [15]. Although photoemission on N doped $TiO_2$ has been extensively studied by several groups [15, 16], the role of oxygen vacancies has not been sufficiently addressed. Here we have studied the electronic structure of well characterized N-doped $TiO_2$ prepared by reactive pulsed laser deposition using HAXPES. The experimental data are compared [17] with theoretical calculations of the density of states (DOS) for $TiO_2$ with N doping both with and without oxygen vacancies. Together, the data and theory demonstrate that N doping produces O vacancies in $TiO_2$. The defect state observed also explains the recent observation of ferromagnetism in non-magnetic doped oxides. The non-magnetic impurity in many cases leads to the formation of oxygen vacancies, which creates a filled Ti *3d* band at the Fermi level and hence the possibility of Stoner splitting of the *3d* band [18].

Experiments were performed at the National Synchrotron Light Source using the National Institute of Standards and Technology beam line X24A. The double-crystal monochromator was operated with Si(111) crystals, and high-resolution photoelectron spectra were obtained with a hemispherical electron analyzer. The photon energy was calibrated prior to each measurement by the Fermi level of a standard Ag foil. Anatase $TiO_2$ with about 7 % N doping was synthesized on Si substrates using reactive pulsed laser deposition. The targets were ablated using an excimer laser ($\lambda$=248 nm) at a constant fluence of 1.8 J/cm$^2$. The base pressure in the chamber was ~$10^{-7}$ Torr. Immediately prior to the deposition, the chamber was backfilled with a mixture of high purity (99.9%) nitrogen and oxygen gas. Details of the synthesis are described elsewhere [14, 18].

Figure 1a shows the Ti *2p* core level for both pure and N doped TiO$_2$ plotted versus electron kinetic energy. The peaks labeled 1 and 2 are the *2p$_{3/2}$* and *2p$_{1/2}$* spin orbit split contributions to the Ti 2p core line from Ti$^{4+}$ [18]. In the case of the N doped sample, we observe two extra peaks labeled 3 and 4. This spectrum is quite similar to that obtained from TiO$_2$ reduced by Ar$^+$ sputtering, and the peak position labeled 3 is very close to the reported values for Ti$^{3+}$ relative to Ti$^{4+}$ [15]. It must be noted that all of our measurements were obtained without Ar$^+$ etch as afforded by the relatively high excitation photon energy. The additional peaks observed in the spectrum have been attributed to Ti$^{3+}$ induced by the formation of N-Ti-O bonds [19], but Nambu et al. [15] have also questioned this conclusion. Based on the relative peak positions of the 3$^{rd}$ and 4$^{th}$ peak we conclude that the N doping of the TiO$_2$ produces Ti$^{3+}$ ions.

The valence-band spectra recorded at photon energy *hv*=2178 eV is shown in figure 1b for both pure and N doped TiO$_2$. The positions of the curve were adjusted by aligning the Ti$^{4+}$ *2p* core level. This allows us to quantitatively compare the change in band gap reported by others to the change in valence band, presuming the conduction band is unaffected by N doping [11]. Due to its formal Ti$^{4+}$ oxidation state, pure TiO$_2$ primarily has a filled O *2p* derived valence band separated from an empty Ti *3d, 4s,* and *4p* derived conduction band by a bulk bandgap of 3.2 eV [20]. The valence-band spectra show the emission from the O *2p* band, and visual examination reveals two features in the N doped sample: a tailing of the valence-band maximum to higher kinetic energy, and an impurity state just above this maximum compared to undoped $TiO_2$. The tail like state is attributed to the N *2p* level since the binding energy of N *2p* is less than O *2p* thus extending the valence-band maximum to lower binding energy. However, the shift observed in the valence band (of about 0.3 eV) is not enough to explain the observed red shift observed in optical absorption spectra reported (of about 0.5-0.8 eV) by several groups [10, 13, 14].

To understand the likely structures produced by N doping of $TiO_2$, we performed density functional theory (DFT) calculations for pure anatase and anatase with various defects, using the VASP package [21]. The calculations used projector augmented wave pseudopotentials [22], with 12 valence electrons for Ti, 6 for O, and 5 for N. The local density approximation (LDA) was used for the exchange-correlation functional. A plane-wave cutoff energy of 353 eV was used with an augmentation charge cutoff of 1500 eV. Density of states calculations were performed on a $\sqrt{2}$ by $\sqrt{2}$ by 1 anatase supercell with 8 Ti and 16 O for pure anatase. Relaxation of cells and atomic positions were performed using 32 k-points in the full Brillouin zone, and density of state (DOS) calculations were then performed using 192 k-points. Because the LDA underestimates the cell volume of anatase, calculations were performed under an artificial negative pressure that reproduces the experimental volume. Four structures were studied: (0,0), (1,0), (0,1), and (1,1), where the first number refers to the number of O->N substitutions and the second number to the number of O vacancies per 24 atom supercell. One N substitution (6.25%) is close to the experimental value. For the (1,1) cell, each possible interdefect geometry was tested. Formation energies of the defects were calculated, with respect to anatase, $O_2$ gas, and $N_2$ gas as "endmember" compounds. The calculated formation energies are (0,1) +6.2 eV; (1,0) +5.5 eV; (1,1) +8.8 eV to +9.7 eV. Importantly, to form an oxygen vacancy in the presence of a nitrogen substitution (+3.3 eV to +4.2 eV, depending on location) is easier than to form an O vacancy in pure $TiO_2$ (+6.2 eV). The most favorable geometries found for the N-substitution/ $Ti^{3+}$/O vacancy complexes are shown in Figure 2. In Fig. 2b, the O vacancy is on the octahedral vertex opposite from the N. In Fig. 2c, two O positions "merge" into one, resulting in an O that is only twofold coordinated with unusually short Ti-O bonds (0.177 to 0.181 nm), and the cell has significantly larger (about 0.01 $nm^3$) volume. Calculations on larger cells confirm that these two geometries have the lowest energies, and the ordering of their energies changes in going from the LDA volume to the experimental volume.

The resulting density of states curves (DOS) convolved with a Gaussian of width 0.4 eV to simulate experiment are shown in Figure 3. The figure shows the calculated DOS for pure, N doped, and N doped plus oxygen vacant $TiO_2$. (Weighted over different geometries by the Boltzmann factor). N substitution leads to a tailing of states near the conduction band edge. Only when O vacancies are also present is there an impurity state at the Fermi energy. We thus conclude on theoretical grounds that N substitution leads to O vacancies, and this conclusion is supported experimentally by the excellent agreement with the DOS curves which will now be demonstrated.

To compare theory and experiment, we have followed the discussion of reference [17] and constructed the theoretical photoemission DOS curves as the cross-section weighted sum of the angular momentum resolved partial density of states of each atom of the unit cell (see figure 4a and 4b). This process accounts for the presence of the N dopants, the formation of O vacancies, and the chemical hybridization in the solid-state electronic structure [17, 23-25]. As in Ref. 17, the cross sections were taken from Ref. 26. Figure 5a and 5b show the comparison between the theoretical and experimental curve for pure and N doped $TiO_2$. Curves are plotted setting the Fermi level to be equal to 0. (Experimental Fermi level was obtained by collecting the VB spectra of a standard Ag foil.) The defect level near the Ti *3d* edge is confirmed as an occupied Ti *3d* level both experimentally and theoretically.

The formation of O vacancy related defect level also explains the knee formation in the optical absorbance spectra of N doped $TiO_2$ reported by several authors [10, 13, 14]. The absorbance tail close to 500 nm [13, 14] could be due to the acceptor level above the valence-band maximum. The experimentally measured VB shows vacancy related defect level at about 1 eV above the valence-band maximum. This agrees well with some of the reported band gap reductions as a result of high concentration of N doping in $TiO_2$. Earlier UPS measurements on $Ar^+$ sputtered $TiO_2$ have shown the position and amplitude of this defect state depend on the amount of the oxygen vacancies or surface reduction [20]. From the foregoing discussions we have now demonstrated that these defect levels are responsible for the change in band gap. We note that the theoretical calculations do not quantitatively reproduce the band gap and the exact band gap reduction due to two effects: (1) the LDA bandgap error (2.4 eV computed for pure anatase vs. 3.2 eV experimentally), and (2) relatively poor theoretical treatment of Ti *3d* electron correlations. For the latter, preliminary LDA+U calculations show better localization of the occupied Ti *3d* states and a lowering of the energies of these states from near the conduction-band minimum to near the valence band maximum as U increases from 0 eV to 8 eV.

In conclusion, we have shown that N doping in anatase $TiO_2$ leads to the formation of oxygen vacancies. The electronic structure is modified, and we directly observe the occupied Ti *3d* states near the Fermi level. The valence-band structure shows tailing to lower binding energies due to the incorporation of less tightly bound N *2p* level which is hybridized with O *2p* level causing a reduction in band gap. However, the shift is not large enough to account for the change in band gap reported from earlier work. Thus the combination of N impurity and the O vacancy is responsible for the UV-vis absorption response in the visible region as well as the visible catalytic activity of N doped $TiO_2$.

A.K.R would like to acknowledge P.D. Siddons (NSLS), Brookhaven National Laboratory for support for this work. This work was performed at the National Synchrotron Light Source, which is supported by the U.S. Department of Energy.

List of Figures:

Figure 1a (Color online): Ti 2p core level for N doped and pure anatase $TiO_2$.

Figure 1b: Valence-band spectra of N doped and pure anatase $TiO_2$. The inset shows the magnified region of the occupied state within the gap.

Figure 2: (a) Structure of pure anatase; (b) N/$Ti^{3+}$/O vacancy defect complex with O opposite N; c) "Merging" of two O into central position. In (b) and (c), the $Ti^{3+}$ ion is obscured from this viewpoint.

Figure 3: Theoretical electronic structures for pure, N doped and N doped plus O vacant $TiO_2$.

Figure 4a: Theoretical angular-momentum resolved components of Ti, O for pure $TiO_2$. (4b) Theoretical angular-momentum resolved components of Ti, O, and N for N doped $TiO_2$ with oxygen vacancies. The curves are normalized with respect to the maximum in each case.

Figure 5a: Theoretical DOS and the experimental valence band for pure anatase TiO$_2$. (5b) Theoretical DOS and the experimental valence band for N doped anatase TiO$_2$. The curves have been scaled to equal peak height.

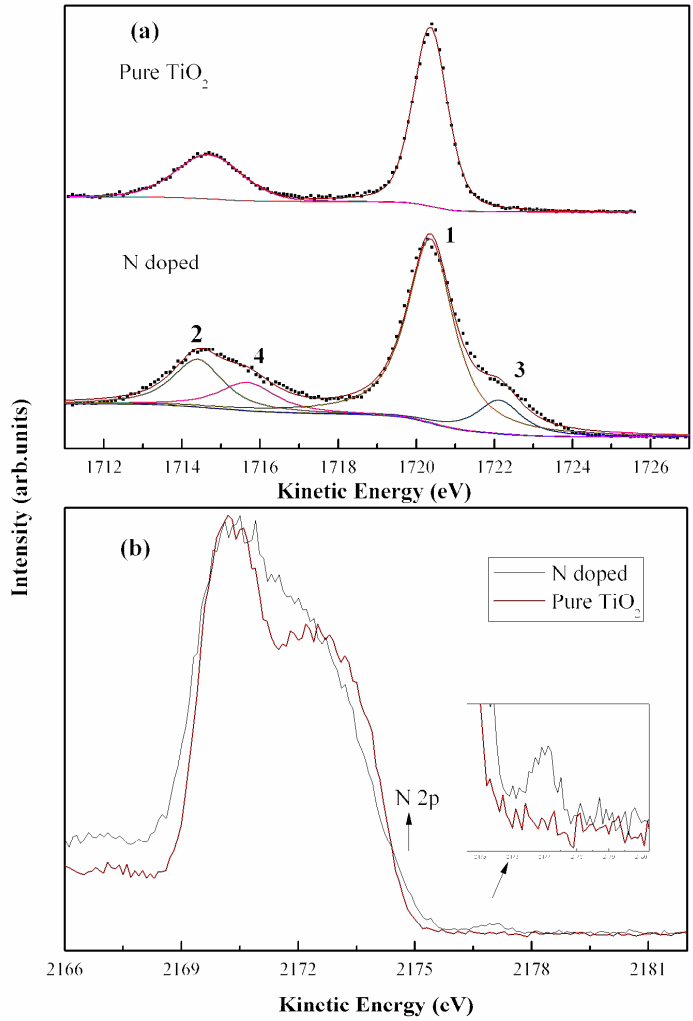

Figure1a&b

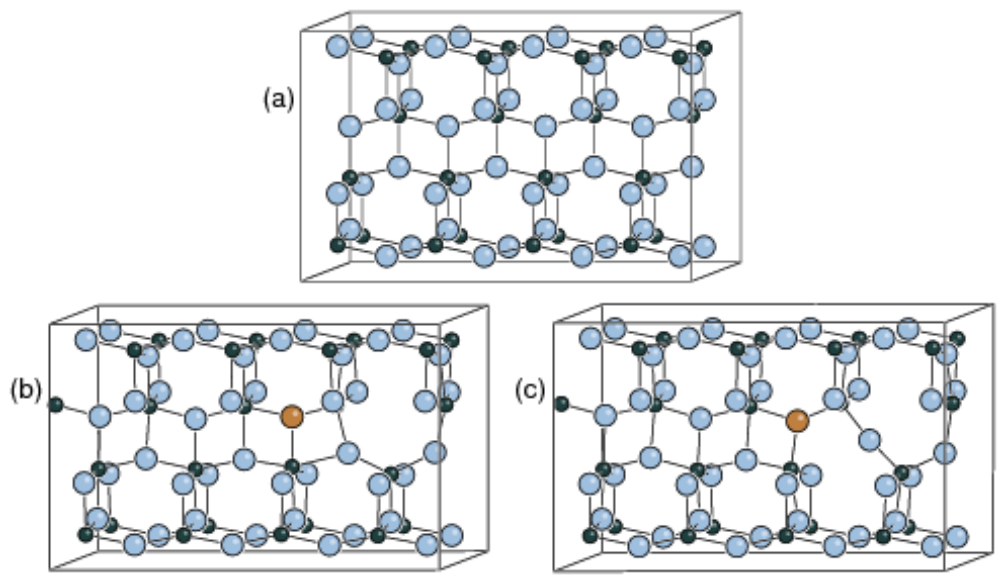

Figure2abc

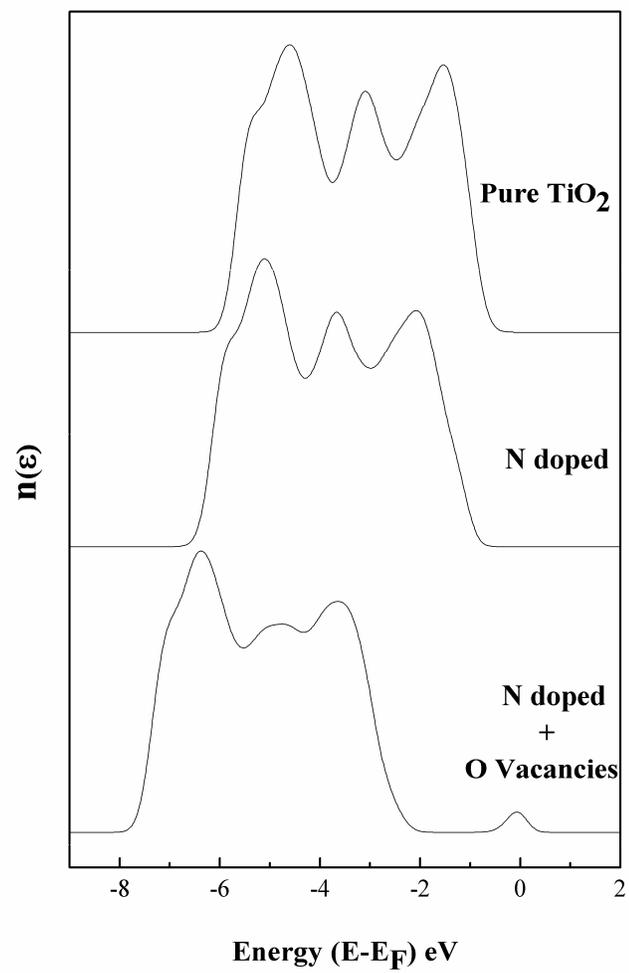

Figure 3

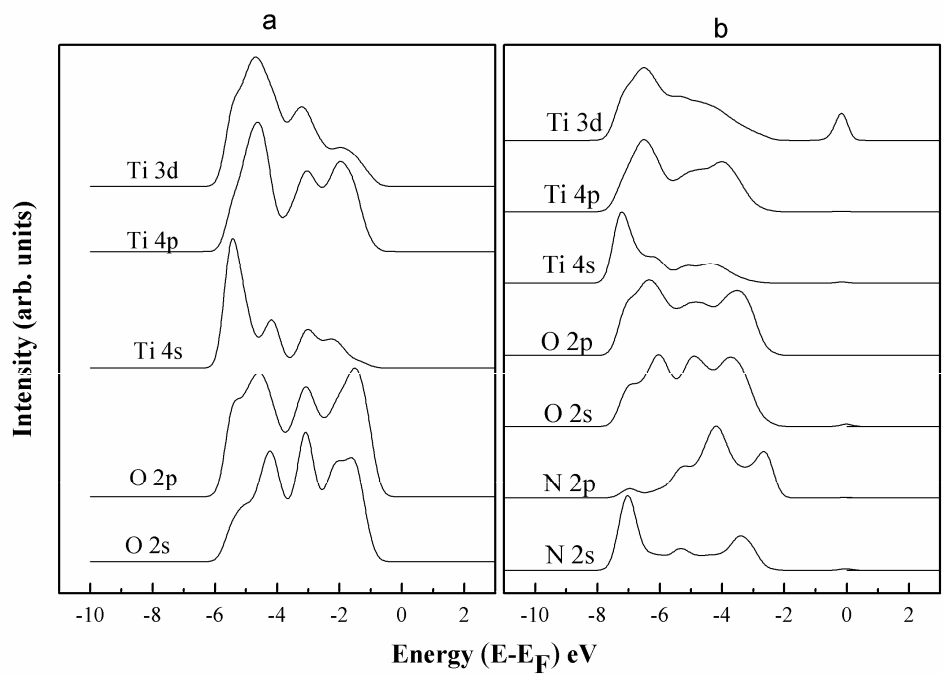

Figure 4ab

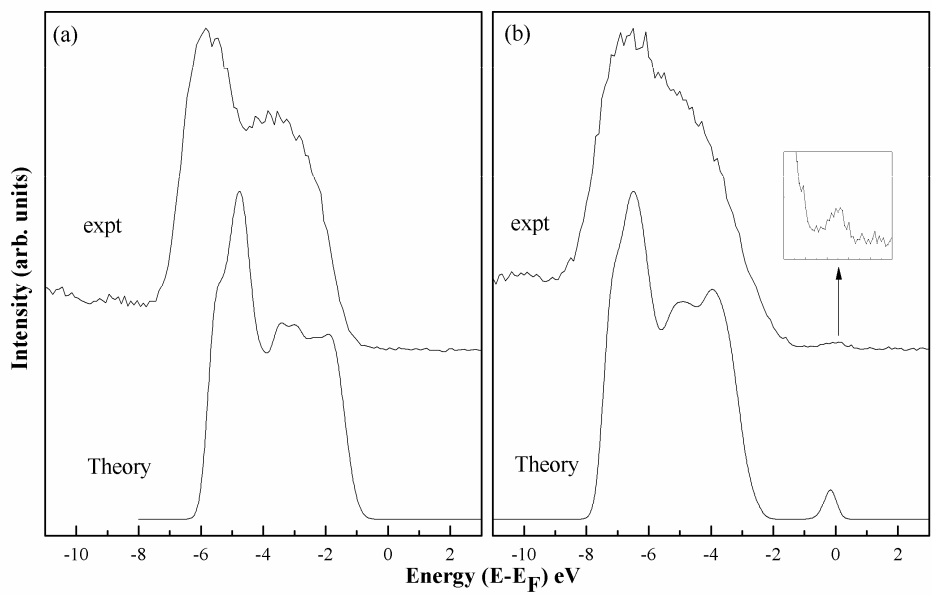

Figure 5ab